\documentclass[a4paper,twoside]{article}

\usepackage{epsfig}
\usepackage{subcaption}
\usepackage{calc}
\usepackage{amssymb}
\usepackage{amsbsy}
\usepackage{amstext}
\usepackage{amsmath}
\usepackage{amsthm}
\usepackage{multicol}
\usepackage{pslatex}
\usepackage{apalike}
\usepackage{algorithm2e}
\usepackage[bottom]{footmisc}
\usepackage{SCITEPRESS}     
\usepackage{graphicx}
\begin{document}

\title{Analysis of Truncated Singular Value Decomposition for Koopman Operator-Based Lane Change Model}

\author{\authorname{Chinnawut Nantabut\sup{1}\orcidAuthor{0000-0002-5767-6023}}
\affiliation{\sup{1}The
Sirindhorn International Thai-German Graduate School of Engineering, King Mongkut's University of Technology North Bangkok, 1518 Pracharat 1 Road, Wongsawang, Bangsue, Bangkok, Thailand}
\email{chinnawut.n@tggs.kmutnb.ac.th}
}


\abstract{Understanding and modeling complex dynamic systems 
is crucial for enhancing vehicle performance and safety, 
especially in the context of autonomous driving. 
Recently, popular methods such as Koopman operators and their approximators, 
known as Extended Dynamic Mode Decomposition (EDMD), 
have emerged for their effectiveness in transforming 
strongly nonlinear system behavior into linear representations. 
This allows them to be integrated with conventional linear controllers. 
To achieve this, Singular Value Decomposition (SVD), 
specifically truncated SVD, is employed to 
approximate Koopman operators from extensive datasets efficiently. 
This study evaluates different basis functions used in 
EDMD and ranks for truncated SVD for representing lane 
change behavior models, aiming to balance computational
efficiency with information loss. The findings, however, 
suggest that the technique of truncated SVD does not necessarily achieve 
substantial reductions in computational training time 
and results in significant information loss.}

\onecolumn \maketitle \normalsize \setcounter{footnote}{0} \vfill

\section{\uppercase{Introduction}}
\label{sec:introduction}

In automotive engineering, model or system identification 
is crucial for understanding the behavior of traffic participants 
under various driving conditions. 
This understanding is essential for improving vehicle performance and safety, 
particularly when transferring insights from scene analysis in these scenarios 
to autonomous vehicles, enhancing their decision-making accuracy.

Given the complexity of modeling such behaviors, 
the underlying dynamic systems are inherently nonlinear. 
To address these challenges, various modeling techniques are employed, 
particularly black-box models, which do not require explicit logical or physical 
representations of the relationship between inputs and outputs. 
Examples of these data-driven approaches include neural networks, 
series models, and autoregressive models \cite{Mauroy20}.

One of the prominent data-driven approaches is the Koopman operator. 
Although initially developed long ago \cite{Koopman31}, 
it has gained renewed attention in modern applications 
for capturing the behavior of dynamical systems. 
Its appeal lies in its connection to classical methods, 
the ability to integrate measurement-based formulations suitable for machine learning, 
and the potential for simplification in real-world applications \cite{Brunton21}. 
Another advantage is that it provides a global representation of the system, 
unlike instantaneous linearization or dynamic linearization, 
which offer only a local approximation of the model around a specific operating point.

Since Koopman operators are theoretically infinite-dimensional, 
the Extended Dynamic Mode Decomposition (EDMD) method is employed to approximate 
them using a set of smaller, more manageable basis functions.

Koopman operators have been widely applied to model identification, 
particularly within the automotive industry \cite{Manzoor23}. 
Notable examples include \cite{Cibulka19}, 
where a single-track model derived from a twin-track model 
was used to generate trajectory data by varying tire forces and kinematic variables.
Their findings showed that increased model complexity does not necessarily improve accuracy. 
Similarly, \cite{Yu22} assessed model fidelity in lane-change scenarios using various systems, 
while \cite{Yu222} demonstrated the ability of Koopman operators to reduce computational 
complexities in vehicle tracking across different road types. 
Moreover, \cite{Kim22} applied Koopman operators to model lane-keeping systems by defining 
states based on lateral dynamics and tire models, utilizing a Linear Quadratic Regulator (LQR) 
for control. Their follow-up work \cite{Kim23} compared multiple basis functions for model 
identification, introducing systematic selection methods and signal normalization techniques. 
Meanwhile, \cite{Joglekar23} investigated function approximators for Koopman operators in 
path-tracking applications.

The operators, particularly when paired with model predictive control (MPC), 
have gained traction for transforming nonlinear models into linear forms, 
enhancing controller performance as shown by \cite{Abraham17}. 
\cite{Gupta22} addressed eco-driving for electric 
vehicles using the Koopman operator with simulator data for high-performance MPC. 
Similarly, \cite{Buzhardt22} generated trajectories using a half-car and 
a terramechanics model in deformable off-road scenarios, applying the 
Koopman operator to lift states like vertical displacement and 
using MPC for regulation. 

Recent studies highlight the increasing use of neural networks as function approximators 
for the Koopman operator. For example, \cite{Han20} demonstrated the integration of 
Koopman operators with MPC, contrasting it with reinforcement 
learning approaches (RL). Similarly, \cite{Xiao23} enhanced long-term predictability by combining 
deep neural networks with Koopman operators, utilizing real driving data. 
In another study, \cite{Guo23} focused on lane-changing scenarios with shared control 
systems, forming the Koopman operator through neural networks and comparing the resulting 
paths to real driver trajectories. Additionally, \cite{Chen24} employed neural networks to 
reformulate states into input constraints for MPC, while \cite{Bongiovanni24} applied the 
Koopman operator to model the behavior of electrical throttle valves. 
Innovatively, \cite{Chen242} used neural network-based autoencoders for approximating 
ldriving behavior and implemented adaptive MPC to handle parameter variations, 
showcasing the growing synergy between neural networks and Koopman operators 
in automotive applications.

As mentioned earlier, Koopman operators enable a 
linear representation of dynamic systems, 
which can be framed in a linear state-space 
formulation using a system matrix. To identify 
this matrix from the available data, it is necessary 
to invert a matrix constructed from these data values. 
Given that this matrix is typically large and asymmetric, 
Singular Value Decomposition (SVD) is utilized to enhance numerical stability. 
Additionally, due to the substantial size of the data, 
the approximation technique known as truncated SVD is employed to reduce the 
dimensions of the matrices involved in the SVD process while preserving as 
much essential information as possible from the original inverted matrix. 
These methodologies not only expedite the training process required to 
derive the system matrix but also come with the trade-off of potential information loss.

Procedures like truncated SVD have not been thoroughly 
explored in the literature, with similar investigations 
primarily found in \cite{Dan23}, which aimed to tackle the overfitting problem. 
However, this study did not address the systematic approximation rank of the matrix, 
and its application was outside the realm of automotive contexts, concentrating 
instead on simple dynamic models. This gap highlights the need for further 
research into the utilization of truncated SVD specifically within automotive applications, 
where the complexities of dynamic systems demand more robust identification techniques.

In this paper, the use of truncated SVD 
for approximating the system matrix derived from the basis functions 
utilized in EDMD is investigated, as discussed in Section \ref{sectionMethod}. 
The focus is centered on a use case involving lane-changing behavior. 
Initially, data is generated using simple geometric models, followed 
by the introduction and selection of basis functions that serve as approximators.
The process of calculating system dynamics through SVD is detailed, 
highlighting the significance of the truncated version and the 
procedure for selecting the appropriate matrix rank. 
All implementations are conducted in Python, 
as outlined in Section \ref{sectionSimulation}. 
Finally, the resulting approximations are analyzed and discussed 
in comparison to the original matrix, considering aspects such as 
information loss and time complexity.

It is important to note that topics like MPC as a controller
are not within the scope of this work and are therefore excluded. 
Unlike the majority of the literature reviewed, the primary focus is 
on system or model identification through the application of Koopman operators, 
alongside the associated techniques of EDMD and truncated SVD.

\begin{figure}[!h]
   \flushleft
   \includegraphics[scale=0.22]{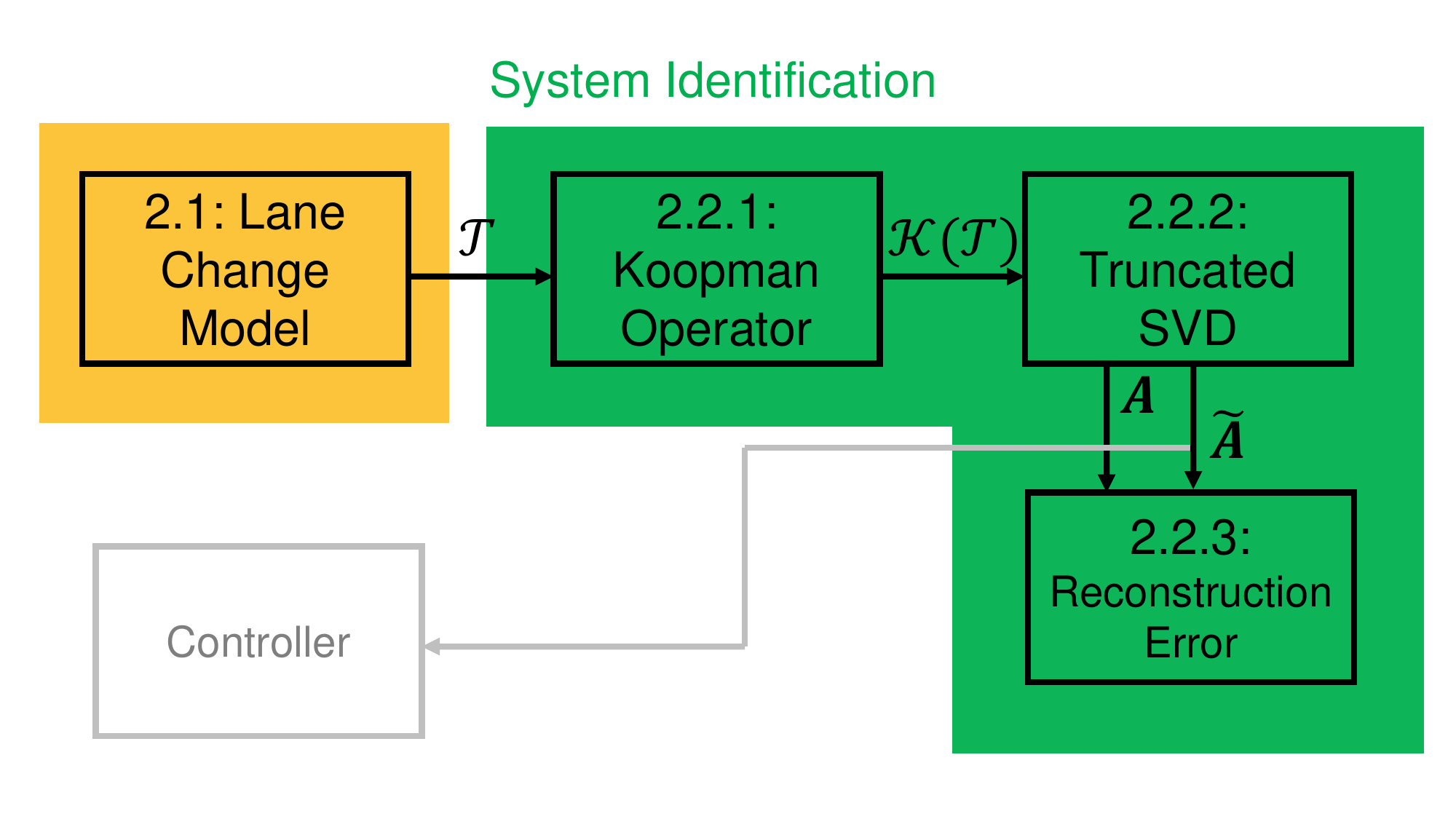}
   \caption{Analysis steps for the truncated SVD-based Koopman operator in a lane change model.}
   \label{fig02}
\end{figure}

\section{METHOD}
\label{sectionMethod}

The analysis presented in this paper is summarized in 
Figure \ref{fig02}. First, a simplified vehicle model 
for lane change behavior is explained in Section \ref{sectionA}, 
leading to the generation of trajectories denoted as $\mathcal{T}$.
Next, the primary focus of this paper - system identification - is executed as detailed 
in Section \ref{sectionB}, which consists of three steps. 

The Koopman operator $\mathcal{K}$ and the EDMD, 
along with their basis functions $\Phi$, are introduced in Section \ref{sectionC}. 
The transformed trajectories $\mathcal{K}(\mathcal{T})$ 
or their approximators $\Phi(\mathcal{T})$ 
are then used to determine the system matrix $\pmb{A}$, 
which describes the linear system.

Subsequently, an appropriate truncated SVD, selected based on specific criteria, 
is discussed in Section \ref{sectionD}, 
resulting in the approximated system matrix $\pmb{\tilde{A}}$. 
While the model for these trajectories is typically employed for controller design, 
this aspect is not addressed in the current paper. Finally, the original matrix $\pmb{A}$ 
and the approximated one $\pmb{\tilde{A}}$ 
are compared to form the reconstruction errors, 
as discussed in Section \ref{sectionE}.

\subsection{Lane Change Model}
\label{sectionA}

\begin{figure}[!h]
   \flushleft
   \includegraphics[scale=0.23]{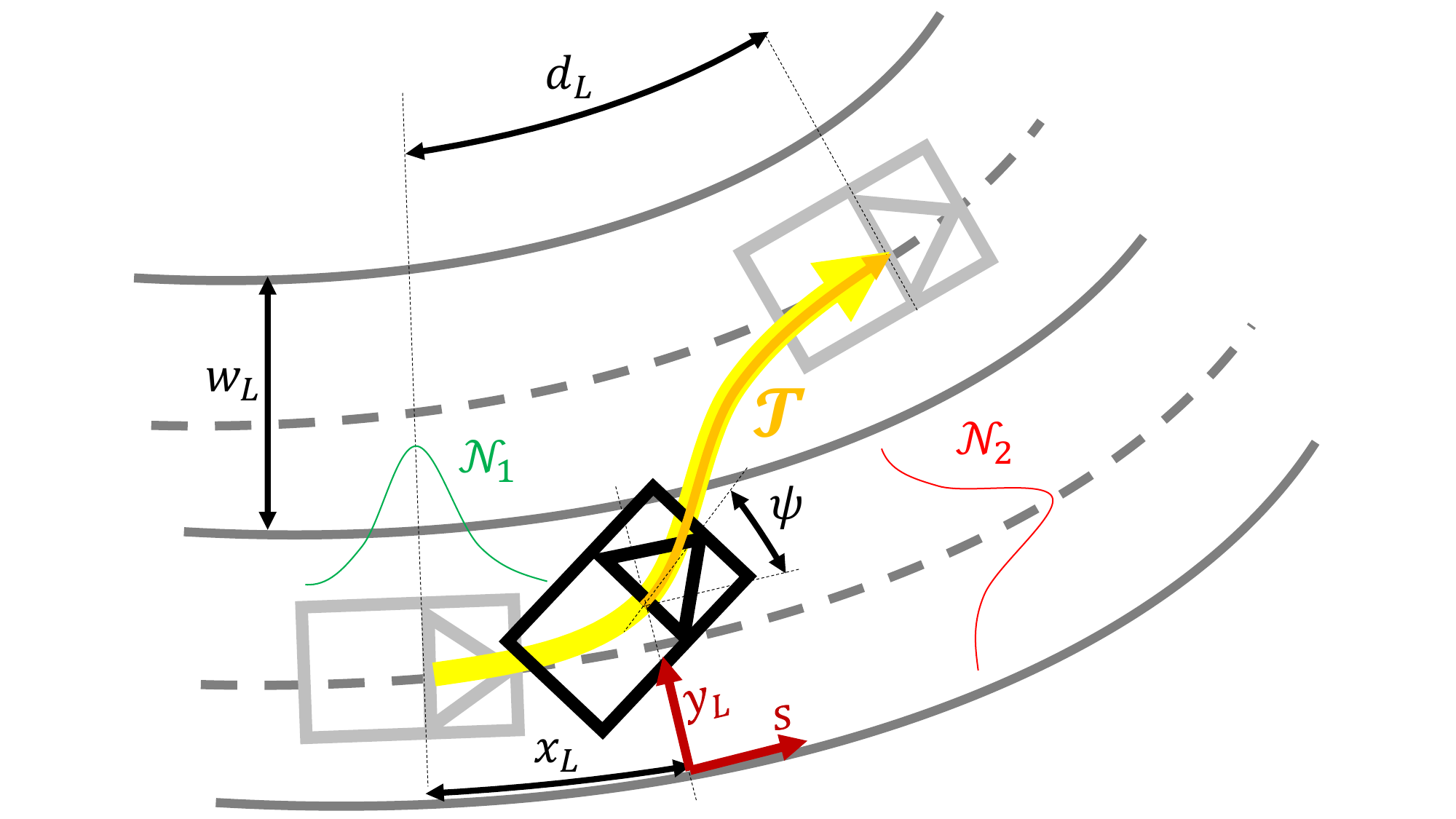}
   \caption{The calculation of lane change trajectories $\mathcal{T}$ is based on \cite{Schreier17}.}
   \label{fig03}
\end{figure}

Since, in this paper, real data were not collected to approximate the model described 
by the Koopman operator, a mathematically simplified model proposed 
by \cite{Schreier17} is used instead, as depicted in Figure \ref{fig03}. 

Additionally, the longitudinal and lateral movement of the vehicle can be modeled independently. 
The vehicle transitions from the right to the left lane of width $w_L$ in the Frenet-Serret frame, 
also known as the lane frame, discribed by the coordinates $s$ and $y_L$. 
The $s$ axis runs along the right edge of the right lane, 
whereas the $y_L$ axis is perpendicular to it and directed to the left.  

Furthermore, the vehicle is located at the position $(x_L, y_L)$ 
and is oriented to the lane at an angle $\psi$. 
The longitudinal length of the complete trajectory, depicted in yellow, is defined by $d_L$. 
Starting from the current pose, including the vehicle's position and orientation, 
a trajectory $\mathcal{T}$ in orange is generated until the vehicle reaches 
the middle line of the left lane.   

To model the longitudinal movement, 
a constant acceleration model is utilized.
In this model, the longitudinal state at time step $k$ is defined as 
$\pmb{\mathcal{X}_{s,k}} = [s_{k} \ v_{s,k} \ a_{s,k}]^T$, where $s_{k}$ 
denotes the longitudinal displacement along the road,
$v_{s,k}$ represents the longitudinal velocity, and $a_{s,k}$ indicates 
the longitudinal acceleration at time step $k$.
The next state $\pmb{\mathcal{X}_{s,k+1}}$ 
can be sampled from the normal distribution $\mathcal{N}_1$,
as shown in green in Figure \ref{fig03}:

\begin{equation}
   \begin{aligned}
      \label{eqt2}
      \begin{bmatrix}
         s_{k+1} \\
         v_{s,k+1} \\
         a_{s,k+1}
      \end{bmatrix}
      \sim
      \mathcal{N} \Biggl(
         \begin{bmatrix}
            1 & T & \frac{1}{2}T^2 \\
            0 & 1 & T \\  
            0 & 0 & 1
           \end{bmatrix}
         \begin{bmatrix}
            s_{k} \\
            v_{s,k} \\
            a_{s,k}
           \end{bmatrix}
           ,
           \\
           \begin{bmatrix}
            \frac{1}{4}T^4 & \frac{1}{2}T^3 & \frac{1}{2}T^2 \\
            \frac{1}{2}T^3 & T^2 & T \\  
            \frac{1}{2}T^2 & T & 1
           \end{bmatrix}\sigma_{a_s}^2
     \Biggr),
   \end{aligned} 
\end{equation}
where the first argument of the expression denotes the mean vector, 
and the second one is represented by the covariance matrix 
that depends on the sample time $T$ and the standard deviation 
in the longitudinal acceleration $\sigma_{a_s}$. 

To model the lateral movement, which is approximated by sinusoidal geometry,
the initial lateral position $y_{L,0}$ is generated
from the normal distribution $\mathcal{N}_2$, as shown in red in Figure \ref{fig03}:

\begin{equation}
   \begin{aligned}
      \label{eqt3}
      y_{L,0}
      \sim
      \mathcal{N} (0.5w_L, \sigma_{y_L}^2),
   \end{aligned} 
\end{equation}
where $\sigma_{y_L}$ denotes the standard deviation in the lateral displacement.

The start angle $\psi_0$ can be sampled from a uniform distribution $\mathcal{U}$, 
with the exclusion of zero:
\begin{equation}
   \begin{aligned}
      \label{eqt4}
      \psi_0
      \sim
      \mathcal{U}]0, \psi_{0,max}],
   \end{aligned} 
\end{equation}
where $\psi_{0,max}$ is the maximal possible initial yaw angle.

As derived in \cite{Schreier17}, where $y_{L,0}$ must also be larger than $0.5w_L$,
the longitudinal length of the complete trajectory 
$d_L$ can then be calculated by:

\begin{equation}
   \begin{aligned}
      \label{eqt5}
      d_L
      = 
      \frac{w_L\pi}{2\tan(\psi_{0})}\cos\Biggl(\sin^{-1}\Biggl(\frac{2y_{L,0}}{w_L}-2\Biggr)\Biggr),
   \end{aligned} 
\end{equation}
as well as the current longitudinal displacement $x_L$: 
\begin{equation}
   \begin{aligned}
      \label{eqt6}
      x_L
      = 
      \Biggl(\frac{1}{2}+\frac{1}{\pi}\sin^{-1}\Biggl(\frac{2y_{L,0}}{w_L}-2\Biggr)\Biggr)d_L.
   \end{aligned} 
\end{equation}

The initial position is then defined by $(x_L, y_{L,0})$, 
where the longitudinal position at time step $k = 0$ is set to $s_{0} = x_L$. 
The longitudinal position $s_{k}$ is updated using the constant acceleration model. 
The new lateral position $y_{L,k}$ is then computed by:

\begin{equation}
   \begin{aligned}
      \label{eqt7}
      y_{L,k}
      = 
      \frac{w_L}{2}\sin\Biggl(\frac{\pi}{d_L}(s_k+x_L)-\frac{\pi}{2}\Biggr)+w_L.
   \end{aligned} 
\end{equation}

Subsequently, the trajectories for the 
following system identification $\mathcal{T}$ 
can be defined by a set of $N_T$ trajectories 
$\mathcal{T}_i$, denoted by:  
\begin{equation}
   \begin{aligned}
      \label{eqt8}
      \mathcal{T}_i
      = 
      \bigcup_{k=0}^{k_{max}}\{(s_k,y_{L,k})_i\}, 
   \end{aligned} 
\end{equation}
where $k_{max} = \max_{k}\{k \| (s_k+x_L) \le d_L\}$.

\subsection{System Identification}
\label{sectionB}
Once the trajectories are generated in Section \ref{sectionA}, 
the next step is to understand how the system evolves based on the presented data. 
Given the assumption that the system is complex and nonlinear, 
Koopman operators and EDMD are employed to elevate the system 
into a higher-dimensional space where it can be described linearly. 
This transformation enables the application of typical linear 
modeling techniques such as LQR 
or linear MPC.

Section \ref{sectionC} elaborates on the Koopman operator 
and its approximators, EDMD, highlighting their roles in system transformation. 
Subsequently, to optimize computational efficiency, 
Section \ref{sectionD} discusses the reduction of linear system dynamics 
using truncated SVD, 
focusing on methods to select an appropriate threshold.

Finally, Section \ref{sectionE} covers the methods for reconstructing 
truncated trajectories and compares them with the original trajectories.

\subsubsection{Koopman Operator}
\label{sectionC}

\begin{figure}[!h]
   \flushleft
   \includegraphics[scale=0.25]{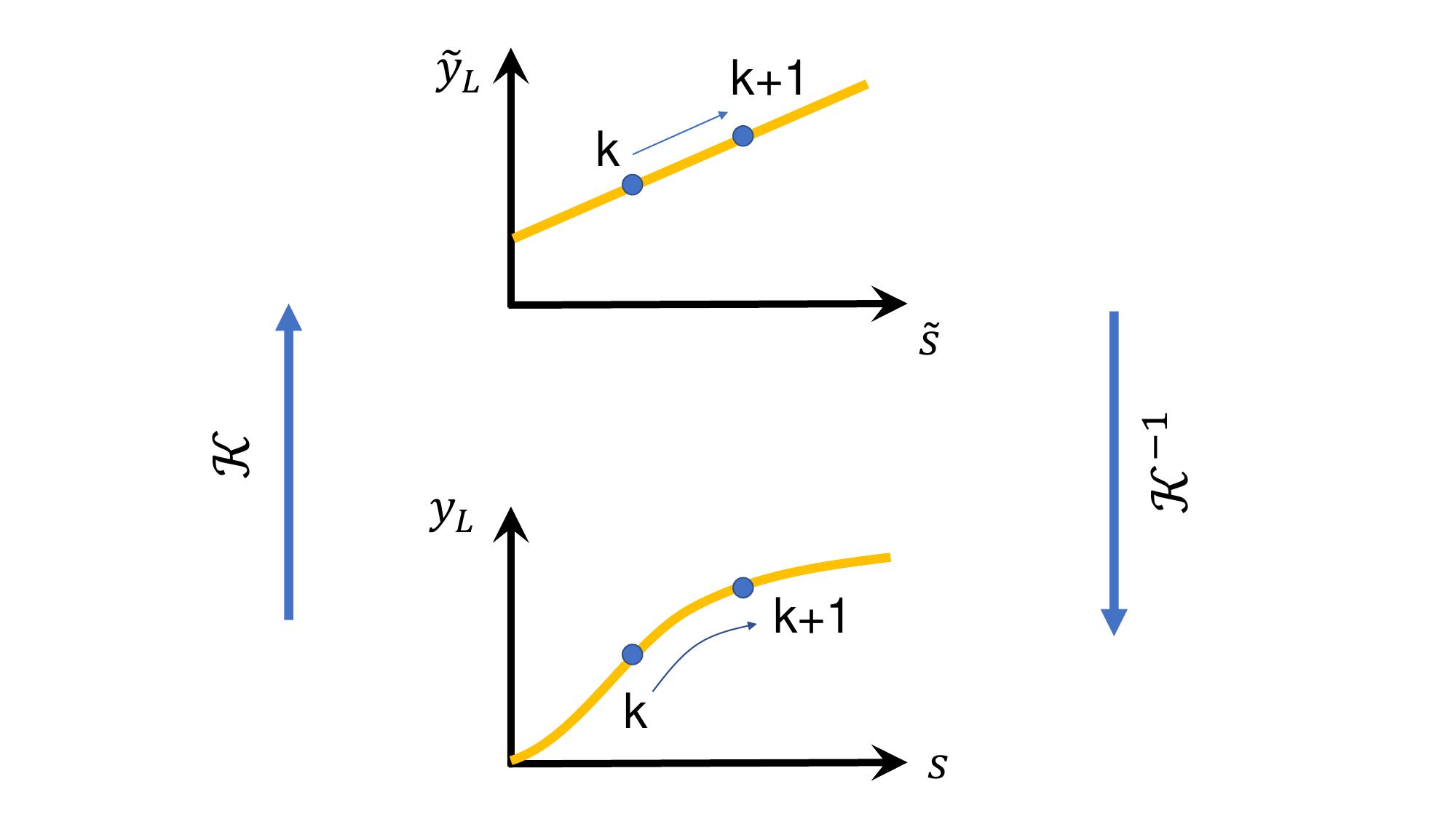}
   \caption{The Koopman operator $\mathcal{K}$ 
   "lifts" states at different time steps, 
   from $(s, y_L)$ to $(\tilde{s}, \tilde{y}_L)$.
   This transformation converts the nonlinear dependency 
   among adjacent states into a linear one. 
   Dealing with the lifted states becomes 
   more manageable compared to the original states. 
}
   \label{fig04}
\end{figure}

The concept of the Koopman operator $\mathcal{K}$ is to transform 
or "lift" the original nonlinearly dependent adjacent states to 
another space where the relationship between the new or lifted 
states becomes linear, as illustrated in Figure \ref{fig04}.
Here, the original states are described by trajectories 
obtained from the previous section, consisting of tuples 
representing longitudinal displacement $s$ and the lateral displacement $y_L$.
Since these trajectories are generated from sinusoidal curves, 
the transition from one state to another is inherently nonlinear.
With the Koopman operator $\mathcal{K}$, 
the states can be transformed into the lifted ones $\mathcal{K}(s,y_L)$ such that 
\begin{equation}
   \begin{aligned}
      \label{eqtK1}
      \mathcal{K}(s,y_L)_{k+1} 
      = 
      \pmb{A}\mathcal{K}(s,y_L)_{k}, 
   \end{aligned} 
\end{equation}
where $\pmb{A}$ is the system matrix describing the linear dependency 
between the current (at time step $k$) lifted states 
$(\tilde{s},\tilde{y}_L)_{k} = \mathcal{K}(s,y_L)_{k} $, 
and the next lifted states 
$(\tilde{s},\tilde{y}_L)_{k+1} = \mathcal{K}(s,y_L)_{k+1}$.

Since the Koopman operator is infinite-dimensional, 
it can only be approximated using EDMD. 
\eqref{eqtK1} can then be transformed into:

\begin{equation}
   \begin{aligned}
      \label{eqtK2}
      \Phi(s,y_L)_{k+1} 
      \approx
      \pmb{A}\Phi(s,y_L)_{k}, 
   \end{aligned} 
\end{equation}
where $\Phi$  denotes the function approximator 
that aims to describe the linearity as accurately as possible.

There are various options for selecting function approximators. 
A common approach involves using a set of basis functions to construct them. 
One widely used type is the monomial basis:

\begin{equation}
   \begin{aligned}
      \label{eqtK3}
      \Phi_m(s,y_L) 
      =
      \begin{bmatrix}
         s \\ y_L \\ s^2 \\ y_L^2 \\ \vdots \\ s^{N_m} \\ y_L^{N_m} \\
      \end{bmatrix},
   \end{aligned} 
\end{equation}
where $N_m$ denotes the highest order of polynomials.

Another common type is the thin plate spline radial basis, described by:
\begin{equation}
   \begin{aligned}
      \label{eqtK4}
      \Phi_r(s,y_L) 
      =
      \begin{bmatrix}
         s \\ y_L \\ 
         \lVert s - c_s \rVert^2_2 \log(\lVert s - c_s \rVert _2) \\ 
         \lVert y_L - c_y \rVert^2_2 \log(\lVert y_L - c_y \rVert _2) \\
      \end{bmatrix},
   \end{aligned} 
\end{equation}
where $\lVert \cdot \rVert _2$ denotes the $l^2$-norm (which, in this context, 
is simply the absolute value), 
and the constants $c_s$ as well as $c_y$ are randomly chosen. 

At the end, the transformed trajectory $\mathcal{K}(\mathcal{T}_i)$ is obtained, 
which can be approximated as:
\begin{equation}
   \begin{aligned}
      \label{eqtK32}
      \Phi(\mathcal{T}_i)
      =
      \bigcup_{k=0}^{k_{max}}\{\Phi(s_k,y_{L,k})_i\}.
   \end{aligned} 
\end{equation}

\subsubsection{Truncated SVD}
\label{sectionD}

In the case of EDMD, once the states are lifted using the operator $\Phi$,
the lifted states are then stored 
and utilized for subsequent training processes. 
To determine the system matrix $\pmb{A}$ 
that captures the linearity, the inversion of 
 $\mathcal{K}(s,y_L)_{k}$ in \eqref{eqtK1} 
or $\Phi(s,y_L)_{k}$ in \eqref{eqtK2} is necessary. 
This involves employing SVD for numerical stability, 
which can also be truncated to expedite the training process. 
However, not only two tuples of $\Phi(s,y_L)_{k}$ are utilized, 
but the entire trajectories are used to determine the system matrix $\pmb{A}$. 
At first, a so-called snapshot matrix of trajectory $\mathcal{T}_i$, 
$\pmb{\mathcal{X}_i}$, can be defined as:
\begin{equation}
   \begin{aligned}
      \label{eqtS1}
      \pmb{\mathcal{X}_i} 
      = \\
      \begin{bmatrix}
         \Phi(s_0,y_{L,0})_i &  \dots  & \Phi(s_{k_{max}-1},y_{L,k_{max}-1})_i\\
        \end{bmatrix}, 
   \end{aligned} 
\end{equation}
and its right-shifted snapshot matrix $\pmb{\mathcal{X}_i'}$ is denoted by: 
\begin{equation}
   \begin{aligned}
      \label{eqtS2}
      \pmb{\mathcal{X}_i'} 
      = \\
      \begin{bmatrix}
         \Phi(s_1,y_{L,1})_i &  \dots & \Phi(s_{k_{max}},y_{L,k_{max}})_i\\
        \end{bmatrix}. 
   \end{aligned} 
\end{equation}

Consequently, the total snapshot matrix of all trajectories $\mathcal{T}$, 
$\pmb{\mathcal{X}}$, is computed by:
\begin{equation}
   \begin{aligned}
      \label{eqtS3}
      \pmb{\mathcal{X}} 
      = 
      \begin{bmatrix}
         \pmb{\mathcal{X}_1} & \pmb{\mathcal{X}_2} & \dots  & \pmb{\mathcal{X}_{N_T}}\\
        \end{bmatrix}, 
   \end{aligned} 
\end{equation}
and its total shifted snapshot matrix $\mathcal{X}'$ is denoted by:  
\begin{equation}
   \begin{aligned}
      \label{eqtS4}
      \pmb{\mathcal{X}'} 
      = 
      \begin{bmatrix}
         \pmb{\mathcal{X}_1'} & \pmb{\mathcal{X}_2'} & ...  & \pmb{\mathcal{X}_{N_T}'}\\
        \end{bmatrix}.
   \end{aligned} 
\end{equation}

The linear dynamics of the system can be approximated, 
using the system matrix $\pmb{A}$, by:
\begin{equation}
   \begin{aligned}
      \label{eqtS5}
      \pmb{\mathcal{X}'} 
      \approx 
      \pmb{A}\pmb{\mathcal{X}}, 
   \end{aligned} 
\end{equation}
analogous to \eqref{eqtK1} and \eqref{eqtK2}. 

Since the matrix $\pmb{\mathcal{X}}$ is not square, 
its pseudo-inverse or Moore-Penrose inverse
$\pmb{\mathcal{X}^{\dagger}}$ 
is computed instead to calculate the system matrix 
$\pmb{A}$: 

\begin{equation}
   \begin{aligned}
      \label{eqtS6}
      \pmb{A} \approx \pmb{\mathcal{X}'\mathcal{X}^{\dagger}}, 
   \end{aligned} 
\end{equation}
where $\pmb{\mathcal{X}^{\dagger}} 
= \pmb{\mathcal{X}^T}(\pmb{\mathcal{X}\mathcal{X}^T})^{-1}$ 
as derived from the so-called normal equation.

Alternatively, the snapshot \textit{real} matrix $\pmb{\mathcal{X}}$ 
can be reformulated using SVD, resulting in the form:

\begin{equation}
   \begin{aligned}
      \label{eqtS7}
      \pmb{\mathcal{X}} = \pmb{U}\pmb{\Sigma}\pmb{V^T},
   \end{aligned} 
\end{equation}
which results in the pseudo-inverse $\pmb{\mathcal{X}^{\dagger}}$ of the form:

\begin{equation}
   \begin{aligned}
      \label{eqtS8}
      \pmb{\mathcal{X}^{\dagger}} = \pmb{V\Sigma^{-1}U^T},
   \end{aligned} 
\end{equation}
where the $\pmb{\Sigma}$-matrix is displayed as: 

\begin{equation}
   \begin{aligned}
      \label{eqtS12}
      \pmb{\Sigma} = 
         \begin{bmatrix}
            \sigma_1 & 0 & 0 & \dots & 0 & 0 &\dots & 0 \\
            0 & \sigma_2 & 0 & \dots & 0 & 0 &\dots & 0\\
            \vdots & \vdots & \vdots & \ddots & \vdots & \vdots &\ddots & \vdots\\
            \vdots & \vdots & \vdots & \dots & \sigma_{r_{max}} & 0 &\dots & 0\\
            \vdots & \vdots & \vdots & \ddots & \vdots & \vdots &\ddots & \vdots \\
            0 & 0 & 0 & \dots & 0 & 0 & \dots & 0 \\
           \end{bmatrix},
   \end{aligned} 
\end{equation}
given that this matrix (and therefore $\pmb{\mathcal{X}}$) has a rank of $r_{max}$, 
the singular values are sorted in descending order, 
namely $\sigma_1 \ge \sigma_2 \ge ... \ge \sigma_{r_{max}}$.

By choosing a specific rank $r \leq r_{max}$, 
an approximated $\pmb{\tilde{\mathcal{X}}}$ in the truncated SVD is obtained by:

\begin{equation}
   \begin{aligned}
      \label{eqtS9}
      \pmb{\tilde{\mathcal{X}}} = \pmb{\tilde{U}\tilde{\Sigma}\tilde{V}^T}, 
   \end{aligned} 
\end{equation}
where $\pmb{\tilde{U}} = \pmb{U}_{:,1:r}$, 
$\pmb{\tilde{\Sigma}} = \pmb{\Sigma}_{1:r,1:r}$, 
and $\pmb{\tilde{V}} = \pmb{V}_{1:r,:}$.

Now, the question of how to systematically choose the rank $r$ is posed. 
As suggested in \cite{Brunton19}, 
the percentual accumulated "energy" 
up to rank $r$ can be calculated as: 

\begin{equation}
   \begin{aligned}
      \label{eqtS13}
      E_r = 100 \% \cdot \frac{\sum_{i=1}^{r}\sigma_i}{\sum_{i=1}^{r_{max}}\sigma_i}.
   \end{aligned} 
\end{equation}

Typically, the empirical values of $E_r = 90 \%$ or $E_r = 99 \%$ 
are used to determine the appropriate rank, 
resulting in a good representation $\pmb{\tilde{\mathcal{X}}}$ 
of the original snapshot matrix $\pmb{\mathcal{X}}$.
Alternatively, a hard threshold (HT) based on the ratio $\beta = \frac{m}{n}$ 
of the dimensions of the (snapshot) matrix 
$\pmb{\mathcal{X}}_{n \times m}$ and the median of its singular values $\sigma_{med}$
is recommended by \cite{Gavish14}.     
The hard threshold $r_{HT}$ is calculated by: 
\begin{equation}
   \begin{aligned}
      \label{eqtS14}
      r_{HT} = \omega(\beta)\sigma_{med},
   \end{aligned} 
\end{equation}
where $\omega(\beta)$ is based purely on the ratio $\beta$ 
and can be calculated accordingly.
Since some values of the 
relationship between $\beta$ and $\omega$ are provided,
interpolation techniques can be employed
to roughly determine the hard threshold rank $r_{HT}$ used to approximate 
the snapshot matrix $\pmb{\mathcal{X}}$.  

The system matrix $\pmb{A}$ can be approximated by 
an approximated system matrix $\pmb{\tilde{A}}$:
\begin{equation}
   \begin{aligned}
      \label{eqtS10}
      \pmb{\tilde{A}} = \pmb{\mathcal{X}'\tilde{V}{\tilde{\Sigma}^{-1}}\tilde{U}^T}.
   \end{aligned} 
\end{equation}

Analogous to \eqref{eqtK2}, the approximated system dynamics can then be described as:
\begin{equation}
   \begin{aligned}
      \label{eqtS11}
      \Phi(s,y_L)_{k+1} \approx \pmb{\tilde{A}}\Phi(s,y_L)_{k}. 
   \end{aligned} 
\end{equation}

The next states $(s,y_L)_{k+1}$ 
can then be retrieved by inverting the operator $\Phi$.

\subsubsection{Reconstruction Error}
\label{sectionE}

To evaluate model fidelity, 
the Frobenius norm $\lVert \pmb{B} \rVert _F$ of a matrix $\pmb{B}$ can be used, 
where $\lVert \pmb{B} \rVert _F = \sqrt{\sum_j\sum_i{b_{ij}^2}}$ 
and $b_{ij}$ denotes the elements of the matrix $\pmb{B}$. 
This norm allows for a comparison of
the system dynamics between the full-ranked system matrix $\pmb{A}$ 
and its truncated version $\pmb{\tilde{A}}$.

The relative reconstruction error 
of the truncated matrix $\pmb{\tilde{A}}$ is represented by:
\begin{equation}
   \begin{aligned}
      \label{eqtR1}
      RE_{\pmb{\tilde{A}}} = \frac{\lVert \pmb{A} - \pmb{\tilde{A}} \rVert _F}{\lVert \pmb{A} \rVert _F}. 
   \end{aligned} 
\end{equation}

The time consumption of the truncated matrix $\pmb{\tilde{A}}$ 
can also be measured relative to the full matrix $\pmb{A}$: 
\begin{equation}
   \begin{aligned}
      \label{eqtR2}
      \tilde{t}_{\pmb{\tilde{A}}} = \frac{t_{\pmb{\tilde{A}}}}{t_{\pmb{A}}}.
   \end{aligned} 
\end{equation}

All the steps outlined in Figure \ref{fig02} are summarized in Algorithm \ref{alg1} 
for the easy implementation of the analysis proposed in this paper.  
This includes the definition of necessary variables, 
the collection of data, the execution of truncated SVD, 
and the analysis of the results using various metrics.

\begin{algorithm}[!h]
  \caption{Truncated SVD for Koopman Operator-Based Lane Change Model}\label{alg1}
  Step I: Define all parameters:
  \begin{enumerate}
   \item Lane Change Model:
   \begin{enumerate}
      \item Road geometry $w_L$
      \item Standard deviations: $\sigma_{a_s}$, and $\sigma_{y_L}$
      \item Time constant $T$
      \item Kinematic variables $a_0$, $v_0$, $s_0$, and $\psi_{0,max}$
   \end{enumerate}
   \item Basis function:
   \begin{enumerate}
      \item Monomial $N_m$
      \item Thin plate spline radial $c_s$ and $c_y$
   \end{enumerate}
  \end{enumerate}

  Step II: Collecting data: Trajectory puffer $\mathcal{T}$ \\
  For each trajectory $i$,  $\mathcal{T}_i \leftarrow \emptyset$ : \\
  \begin{enumerate}
   \item Sample the initial lateral displacement $y_{L,k} = y_{L,0}$ from \eqref{eqt3}
   and the initial yaw angle $\psi_0$ from \eqref{eqt4}
   \item Calculate the trajectory's longitudinal length $d_L$ from \eqref{eqt5} 
   and the current displacement $s_k = x_L$ from \eqref{eqt6}
   \item $\mathcal{T}_i \leftarrow \{ (s_k, y_{L,k}) \}$
   \item While $s_k \leq d_L$: 
      \begin{enumerate}
         \item Update the longitudinal displacement $s_k$ using \eqref{eqt2}
         and the lateral displacement $y_{L,k}$ using \eqref{eqt7}
         \item $\mathcal{T}_i \leftarrow \mathcal{T}_i \cup \{ (s_k, y_{L,k}) \}$
      \end{enumerate}
   \item Update $\mathcal{T} \leftarrow \mathcal{T} \cup \mathcal{T}_i$
   \end{enumerate}

   Step III: Truncated SVD
   \begin{enumerate}
      \item Calculate snapshot matrices $\pmb{\mathcal{X}}$ and $\pmb{\mathcal{X'}}$ 
      based on \eqref{eqtS1}, \eqref{eqtS2}, \eqref{eqtS3} and \eqref{eqtS4} 
      \item Use SVD on the snapshot matric $\pmb{\mathcal{X}}$ using \eqref{eqtS7} 
      \item Calculate the rank $r$ using \eqref{eqtS13} or \eqref{eqtS14}
      \item Construct truncated snapshot matrix $\pmb{\tilde{\mathcal{X}}}$ using \eqref{eqtS9}
      and the system matrix $\pmb{\tilde{A}}$ using \eqref{eqtS10}  
   \end{enumerate}

   Step IV: Evaluation
   \begin{enumerate}
      \item Calculate the relative reconstruction error $RE_{\pmb{\tilde{A}}}$ using \eqref{eqtS6}, \eqref{eqtS10} and \eqref{eqtR1}
      and the relative time consumption $\tilde{t}_{\pmb{\tilde{A}}}$ using \eqref{eqtR2}
   \end{enumerate}

\end{algorithm}

\section{SIMULATION}
\label{sectionSimulation}

All parameters utilized in the previous section, 
as summarized in Algorithm \ref{alg1}, are defined here. 
The comparison between the original data and the Koopman 
operators using different types of basis functions is illustrated. 
Additionally, qualitative trajectories based on truncated SVD with 
various rank selections are compared and analyzed. 
Finally, the time consumption for these processes is also discussed.

\subsection{Setup}

Taken from the original paper for modeling lane change behavior \cite{Schreier17},
the lane width $w_L$ is set to $3.5 \ m$. 
With the vehicle width $w_V$ of $1.5 \ m$, 
the standard deviation of lateral deviation $\sigma_{y_L} = \frac{1}{3}0.5(w_L-w_V) = \frac{1}{3} \ m$,
whereas the value of the longitudinal direction $\sigma_{a_s} = \frac{0.2}{3} \ \frac{m}{s^2}.$
The time constant $T$ is set to $0.1 \ s.$ 
The initial kinematic variables are $s_0 = 0 \ m, v_0 = 10 \ \frac{m}{s}$, and $a_0 = 0 \ \frac{m}{s^2}$.
The vehicle can start with a maximal orientation of $\psi_{0} = 15 ^{\circ}.$ 

To ensure both basis functions have the same dimension, $N_m$ is set to $2$.
Furthermore, the constants $c_s$ and $c_y$ are sampled from a uniform 
distribution ranging between $-\frac{w_L}{2}$ and $+\frac{w_L}{2}$.

\subsection{Results}
\label{sectionResults}

\begin{figure}[!h]
   \centering
    {\epsfig{file = 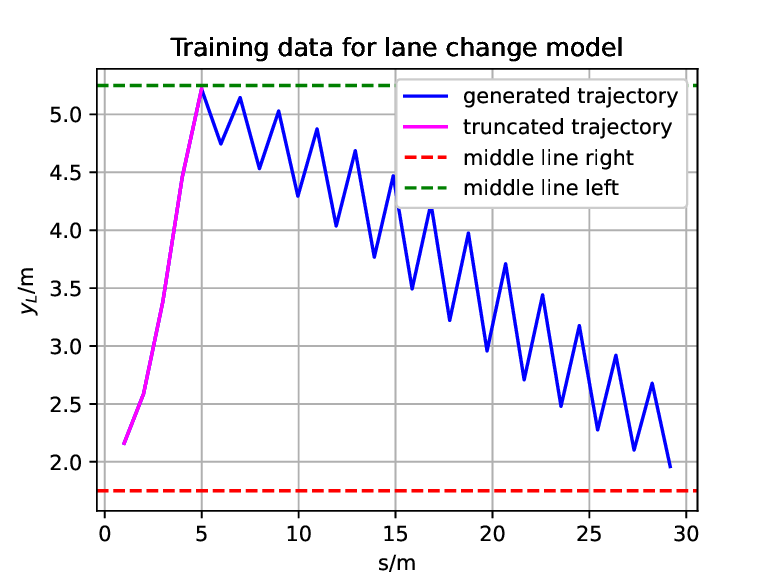, width = 8cm}}
   \caption{An exemplary original trajectory 
   used for training the system matrix $\pmb{A}$ or $\pmb{\tilde{A}}$ 
   illustrates a disadvantage of choosing this model 
   when the longitudinal displacement $(s_k+x_L)$ reaches 
   the maximum longitudinal sinusoidal length $d_L$,  
   as explained in Equation \ref{eqt8}.}
   \label{fig:fig1}
\end{figure}

In Figure \ref{fig:fig1}, a single trajectory illustrating a lane change maneuver 
(from the red right middle lane to the green left middle lane) is depicted in blue, 
based on the approach by \cite{Schreier17} as detailed in Section \ref{sectionA}. 
The vehicle follows a simple sinusoidal pattern, 
returning to the right lane after exceeding the maximum longitudinal displacement $d_L$, 
as previously discussed. 

Subsequently, strong oscillations occur, stemming from the constant acceleration 
model used for longitudinal movement, impacting the lateral movement as well. 
Therefore, the generated data needs truncation before further model training. 
However, this truncation results, as shown in magenta, in shorter trajectories than anticipated, 
potentially leading to model overfitting. 

\begin{figure}[!h]
   \centering
    {\epsfig{file = 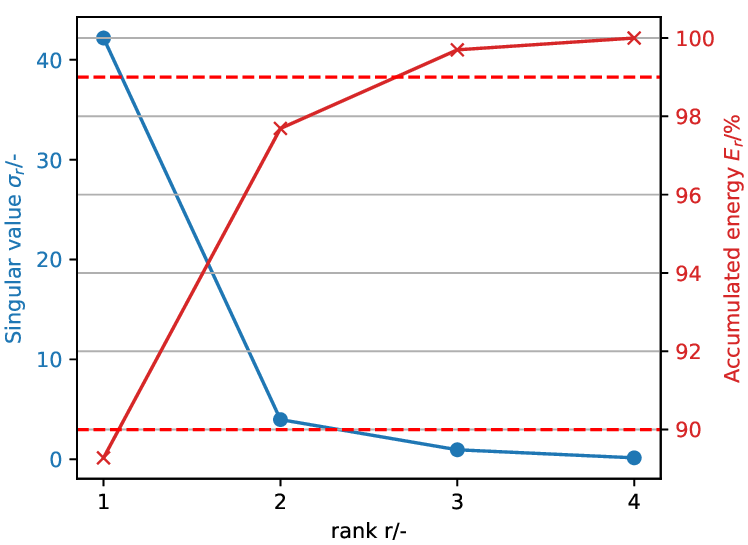, width = 7.5cm}}
   \caption{The sorted singular values $\sigma_r$ and their accumulated energy $E_r$ 
   are visualized against their ranks $r$, using the example of monomial basis functions. 
   Horizontal lines representing the empirical values $E_r = 90 \% $ and $E_r = 99 \% $ 
   are also plotted to aid in identifying their corresponding singular value.}
   \label{fig:fig2}
\end{figure}

Now, the truncated trajectory is subjected to truncated SVD 
by constructing the matrix $\pmb{\mathcal{X}}$ and sorting its singular values $\sigma_i$.
The accumulated energy $E_r$ is calculated for each singular value, and their 
values are plotted in Figure \ref{fig:fig2} using monomial basic functions. 
In this example, it is observed that approximately $E_r = 90 \ \% $ is achieved  
directly at the first singular value ($E_r \approx 89 \%$), as it is relatively large 
($\geq 30$) compared to the others. According to the plot, 
the third singular value corresponds to $E_r = 99 \%$.

However, upon calculating the hard threshold rank $r_{HT}$, 
it is found that it exceeds $4$. Therefore, the full rank is used in this case instead. 
Additionally, similar observations have been made for the radial basis functions, 
resulting in identical rank selections.

\begin{table}[h]
   \caption{Comparison of the truncated SVD performance ($E_r = 90$\%, $E_r = 99$\% and $HT$: hard threshold) 
   based on different basis functions ($m$: monomial and $r$: thin plate spline radial), 
   including metrics such as the relative reconstruction error 
   and the time consumption.}
   \label{tab:example1} \centering
   \begin{tabular}{|c|c|c|c|}
     \hline
     Basis & rank $r$ & $RE_{\pmb{\tilde{A}}}$/\% &  $min\{\tilde{t}_{\pmb{\tilde{A}}}\}/\%$\\
     \hline
     $\Phi_{m, 90\%}$ & 1 & 99.54 & 89.97\\
     \hline
     $\Phi_{m, 99\%}$ & 3 & 95.76  & 96.47\\
     \hline
     $\Phi_{m, HT}$ & 4 (full) & 0 & 100\\
     \hline
     $\Phi_{r, 90\%}$ & 1 & 98.14 & 95.16\\
     \hline
     $\Phi_{r, 99\%}$ & 3 & 60.81 & 97.21\\
     \hline
     $\Phi_{r, HT}$ & 4 (full) & 0 & 100\\
     \hline
   \end{tabular}
\end{table}

The relative reconstruction error $RE_{\pmb{\tilde{A}}}$ 
and the \textit{mininmal} relative time consumption $min\{\tilde{t}_{\pmb{\tilde{A}}}\}$
of the system matrix $\pmb{A}$ are compared in Table \ref{tab:example1}.
It is observed for each type of basis function that the reconstruction error 
decreases with higher rank selections. 

Additionally, in the lane change model with the previously collected data, 
the radial basis function demonstrates superior model fidelity. 
However, with truncated SVD, the reconstruction errors are relatively large ($> \ 50 \%$),
prompting the use of the full-ranked matrix $\pmb{A}$ due to the hard threshold.

Furthermore, truncated SVD does not consistently reduce time consumption, 
as evidenced by cases where $\tilde{t}_{\pmb{\tilde{A}}}$ exceeds $100 \ \% $. 
The \textit{minimal} time consumption is highlighted here, 
underscoring the potential for reduced computation time with truncated SVD.

\section{CONCLUSIONS}

This study investigates the efficacy of truncated SVD in conjunction 
with Koopman operators and EDMD for system identification in the context 
of lane change behavior. The results indicate that while these techniques 
do not necessarily lead to a significant reduction in the computational 
time required for training the system matrix, they entail a compromise in model fidelity. 

To validate these findings, future research will explore the use of diverse datasets 
for training and conduct a more comprehensive statistical analysis, as well as analyze 
controllers not covered here.

\bibliographystyle{apalike}
{\small
\bibliography{example}}

\begin{thebibliography}{}

\bibitem[Abraham et~al., 2017]{Abraham17}
Abraham, I., Torre, G. D.~L., and Murphey, T.~D. (2017).
\newblock Model-based control using koopman operators.
\newblock {\em CoRR}, abs/1709.01568.

\bibitem[Bongiovanni et~al., 2024]{Bongiovanni24}
Bongiovanni, N., Mavkov, B., Martins, R., and Allibert, G. (2024).
\newblock {Data-Driven Nonlinear System Identification of a Throttle Valve
  Using Koopman Representation}.
\newblock In {\em {American Control Conference (ACC 2024)}}, Toronto, Canada.

\bibitem[Brunton et~al., 2021]{Brunton21}
Brunton, S.~L., Budišić, M., Kaiser, E., and Kutz, J.~N. (2021).
\newblock Modern koopman theory for dynamical systems.

\bibitem[Brunton and Kutz, 2019]{Brunton19}
Brunton, S.~L. and Kutz, J.~N. (2019).
\newblock {\em Data-Driven Science and Engineering: Machine Learning, Dynamical
  Systems, and Control}.
\newblock Cambridge University Press, USA, 1st edition.

\bibitem[Buzhardt and Tallapragada, 2022]{Buzhardt22}
Buzhardt, J. and Tallapragada, P. (2022).
\newblock A koopman operator approach for the vertical stabilization of an
  off-road vehicle.
\newblock {\em IFAC-PapersOnLine}, 55(37):675--680.
\newblock 2nd Modeling, Estimation and Control Conference MECC 2022.

\bibitem[Chen et~al., 2024a]{Chen24}
Chen, H., He, X., Cheng, S., and Lv, C. (2024a).
\newblock Deep koopman operator-informed safety command governor for autonomous
  vehicles.
\newblock {\em IEEE/ASME Transactions on Mechatronics}, pages 1--11.

\bibitem[Chen et~al., 2024b]{Chen242}
Chen, Z., Chen, X., Liu, J., Cen, L., and Gui, W. (2024b).
\newblock Learning model predictive control of nonlinear systems with
  time-varying parameters using koopman operator.
\newblock {\em Applied Mathematics and Computation}, 470:128577.

\bibitem[Cibulka et~al., 2019]{Cibulka19}
Cibulka, V., Hanis, T., and Hromcik, M. (2019).
\newblock Data-driven identification of vehicle dynamics using koopman
  operator.
\newblock {\em 2019 22nd International Conference on Process Control (PC19)},
  pages 167--172.

\bibitem[Gavish and Donoho, 2014]{Gavish14}
Gavish, M. and Donoho, D.~L. (2014).
\newblock The optimal hard threshold for singular values is $4/\sqrt {3}$.
\newblock {\em IEEE Transactions on Information Theory}, 60(8):5040--5053.

\bibitem[Guo et~al., 2023]{Guo23}
Guo, W., Zhao, S., Cao, H., Yi, B., and Song, X. (2023).
\newblock Koopman operator-based driver-vehicle dynamic model for shared
  control systems.
\newblock {\em Applied Mathematical Modelling}, 114:423--446.

\bibitem[Gupta et~al., 2022]{Gupta22}
Gupta, S., Shen, D., Karbowski, D., and Rousseau, A. (2022).
\newblock Koopman model predictive control for eco-driving of automated
  vehicles.
\newblock In {\em 2022 American Control Conference (ACC)}, pages 2443--2448.

\bibitem[Han et~al., 2020]{Han20}
Han, Y., Hao, W., and Vaidya, U. (2020).
\newblock Deep learning of koopman representation for control.
\newblock In {\em 2020 59th IEEE Conference on Decision and Control (CDC)},
  pages 1890--1895.

\bibitem[Joglekar et~al., 2023]{Joglekar23}
Joglekar, A., Samak, C., Samak, T., Kosaraju, K.~C., Smereka, J., Brudnak, M.,
  Gorsich, D., Krovi, V., and Vaidya, U. (2023).
\newblock Analytical construction of koopman edmd candidate functions for
  optimal control of ackermann-steered vehicles.
\newblock {\em IFAC-PapersOnLine}, 56(3):619--624.
\newblock 3rd Modeling, Estimation and Control Conference MECC 2023.

\bibitem[Kim et~al., 2022]{Kim22}
Kim, J.~S., Quan, Y.~S., and Chung, C.~C. (2022).
\newblock Data-driven modeling and control for lane keeping system of automated
  driving vehicles: Koopman operator approach.
\newblock In {\em 2022 22nd International Conference on Control, Automation and
  Systems (ICCAS)}, pages 1049--1055.

\bibitem[Kim et~al., 2023]{Kim23}
Kim, J.~S., Quan, Y.~S., and Chung, C.~C. (2023).
\newblock Koopman operator-based model identification and control for automated
  driving vehicle.
\newblock {\em International Journal of Control, Automation and Systems 21},
  page 2431–2443.

\bibitem[Koopman, 1931]{Koopman31}
Koopman, B.~O. (1931).
\newblock Hamiltonian systems and transformation in hilbert space.
\newblock {\em Proceedings of the National Academy of Sciences},
  17(5):315--318.

\bibitem[Manzoor et~al., 2023]{Manzoor23}
Manzoor, W.~A., Rawashdeh, S., and Mohammadi, A. (2023).
\newblock Vehicular applications of koopman operator theory—a survey.
\newblock {\em IEEE Access}, 11:25917--25931.

\bibitem[Mauroy and Goncalves, 2020]{Mauroy20}
Mauroy, A. and Goncalves, J. (2020).
\newblock Koopman-based lifting techniques for nonlinear systems
  identification.
\newblock {\em IEEE Transactions on Automatic Control}, 65(6):2550--2565.

\bibitem[Schreier, 2017]{Schreier17}
Schreier, M. (2017).
\newblock Bayesian environment representation, prediction, and criticality
  assessment for driver assistance systems.
\newblock {\em at - Automatisierungstechnik}, 65(2):151--152.

\bibitem[Wilson, 2023]{Dan23}
Wilson, D. (2023).
\newblock Koopman operator inspired nonlinear system identification.
\newblock {\em SIAM Journal on Applied Dynamical Systems}, 22(2):1445--1471.

\bibitem[Xiao et~al., 2023]{Xiao23}
Xiao, Y., Zhang, X., Xu, X., Liu, X., and Liu, J. (2023).
\newblock Deep neural networks with koopman operators for modeling and control
  of autonomous vehicles.
\newblock {\em IEEE Transactions on Intelligent Vehicles}, 8(1):135--146.

\bibitem[Yu et~al., 2022a]{Yu22}
Yu, S., Shen, C., and Ersal, T. (2022a).
\newblock Autonomous driving using linear model predictive control with a
  koopman operator based bilinear vehicle model.
\newblock {\em IFAC-PapersOnLine}, 55(24):254--259.
\newblock 10th IFAC Symposium on Advances in Automotive Control AAC 2022.

\bibitem[Yu et~al., 2022b]{Yu222}
Yu, S., Sheng, E., Zhang, Y., Li, Y., Chen, H., and Hao, Y. (2022b).
\newblock Efficient nonlinear model predictive control of automated vehicles.
\newblock {\em Mathematics}, 10(21).

\end{thebibliography}

\end{document}